\documentclass[preprint,review,11pt]{elsarticle}

\usepackage{lineno}
  \usepackage{subfigure}
  \usepackage{url}
  \usepackage[normalem]{ulem}  
  \usepackage{subfigure}
  \usepackage{bm}
  \usepackage{multirow}
  \usepackage{graphicx}               %figure insertion with \includegraphics[width=x.cm]{file.eps}
  \usepackage[normalem]{ulem}
  \usepackage{amsmath}
  \usepackage{longtable}
\usepackage[margin=1.1in]{geometry}

\journal{arXiv}

\begin{document}

\begin{frontmatter}

\title{HazeDose: Design and Analysis of a Personal Air Pollution Inhaled Dose Estimation System using Wearable Sensors}

%\author{Ke Hu\fnref{myfootnote}}
\author[mymainaddress]{Ke Hu\corref{mycorrespondingauthor}}
\cortext[mycorrespondingauthor]{Corresponding author: Ke Hu, School of Electrical Engineering and Telecommunications, UNSW, Sydney, NSW 2052, Australia. Tel: +61 4 0579 0900.}
%\address{School of Electrical Engineering and Telecommunications, UNSW, Sydney, NSW 2052, Australia}
\ead{kehuau@gmail.com}

\author[mysecondaryaddress]{Ashfaqur Rahman}
\ead{ashfaqur.rahman@csiro.au}

\author[mymainaddress]{Hassan Habibi Gharakheili}
\ead{h.habibi@unsw.edu.au}

\author[mymainaddress]{Vijay Sivaraman}
\ead{vijay@unsw.edu.au}

\address[mymainaddress]{University of New South Wales, Sydney, Australia}
\address[mysecondaryaddress]{Data61, CSIRO, Sandy Bay, Australia}

\begin{abstract}
Nowadays air pollution becomes one of the biggest world issues in both developing and developed countries. Helping individuals understand their air pollution exposure and health risks, the traditional way is to utilize data from static monitoring stations and estimate air pollution qualities in a large area (usually at suburb or city level) by government agencies. Data from such sensing system is very sparse and cannot reflect real personal exposure. In recent years, several research groups, including ours, have developed participatory air pollution sensing systems which use wearable or vehicle-mounted portable units coupled with smart phones to crowd-source urban air pollution data. These systems have shown remarkable improvement in spatial granularity over government-operated fixed monitoring systems, and lead to better mapping and understanding of urban air pollution. In this paper, we extend the paradigm to HazeDose system, which can personalize the individuals' air pollution exposure. Specifically, we combine the pollution concentrations obtained from an air pollution estimation system -- HazeEst with the activity data from the individual's on-body activity monitors to estimate the personal inhalation dosage of air pollution. Users can visualize their personalized air pollution exposure information via a mobile application. We show that different activities, such as walking, cycling, or driving, impacts their dosage, and commuting patterns contribute to a significant proportion of an individual's daily air pollution dosage. Moreover, we propose a dosage minimization algorithm, with the trial results showing that up to 14.1\% of a biker's daily exposure can be reduced applying our algorithm using fixed routes, while using alternative routes the driver can inhale 25.9\% less than usual. One heuristic algorithm is also introduced to balance the execution time and dosage reduction for alternative routes scenarios. The results show that up to 20.3\% dosage reduction can be achieved when the execution time is almost one seventieth of the original one. The HazeDose system is a step towards enabling accurate medical inferencing of the impact of long-term air pollution on individual health.
\end{abstract}

\begin{keyword}
Wireless sensor network; optimization model; activity pattern; inhaled dosage; mobile application.
\end{keyword}

\end{frontmatter}

% \linenumbers

\section{Introduction}
One of the basic requirements of human health and well-being is clean air. However, air pollution has become one of the world's biggest environmental problems in the past few decades and has had harmful effects on both human health and the environment. The World Health Organization (WHO) estimates that 92\% of the world’s population live in places where air quality levels exceed WHO limits, and outdoor air pollution is associated with approximately three million deaths per year worldwide \cite{WHO}. Indoor air pollution can be just as deadly. In 2012, an estimated 6.5 million deaths (11.6\% of all global deaths) were associated with indoor and outdoor air pollution together. In fact, in Australia, 2,400 deaths per year on average are linked to air quality and health issues, much more than the 1,700 people who die on our roads \cite{CSIRO04}. WHO also reports that in 2012, around seven million people died - one in eight of total global deaths as a result of air pollution exposure.

Most of the current epidemiological evidence on the effects of air pollution is based on assigning estimated exposures using area of residence or workplace. However, a significant amount of commuting time is taken by residents in the major cities such as Sydney surrounded by high-density traffic. Given the spatial and temporal variability in the concentrations of pollutants, short-term individual exposure depends significantly on commuting patterns instead of static residence or working sites. Activity information is also critical to estimate personal exposure. As an example, consider two individuals who are both in the same place at the same time, but one is driving while the other is jogging. They will experience the impact of air pollution in different ways, as they will inhale different amounts due to their various breathing rates, and may additionally have different medical predispositions to the exposure. When these differences are accumulated over an extended period of time, they can become significant, leading to different health outcomes.

In recent years, wearable technologies have become commercially viable and commonplace, which makes personal dose estimation achievable. Therefore, in this paper, we combine ambient pollution levels (taken from our HazeEst system as introduced in \cite{7892954}) with an individual's activity information from wearable sensors to estimate the personal inhalation dosage, which can then be used to make further medical inferences for that individual. Our specific contributions are:

\begin{enumerate}
\item We build HazeDose: a novel system combines location data from mobile phones, human activity data from wearable activity sensor devices, and air pollution data from the HazeEst system to estimate personal air pollution inhaled dosage. 
\item We conduct field trials with our system in Sydney and obtain pollution inhalation dosage estimates showing that commuting is non-negligible in pollution dosage estimation, and it contributes to a large proportion of a person's daily exposure. We also show that different commute patterns (driving, cycling, and walking) entail very different levels of exposure. The estimates from our system are compared to earlier systems that do not include personal activity information, indicating that our system allows for more accurate medical inferencing.
\item We develop a novel dosage minimization algorithm for individuals to manage and reduce their dose. The trial results show that up to 14.1\% of a biker's daily exposure can be reduced applying our algorithm using fixed routes, while using alternative routes the driver can inhale 25.9\% less than usual. One heuristic algorithm is also introduced to balance the execution time and dosage reduction for alternative routes scenarios. The results show that up to 20.3\% dosage reduction can be achieved when the execution time is almost one seventieth of the original one.
\end{enumerate}

%Paper organization
The rest of this paper is organized as follows: \S\ref{sec:related work} discusses prior work relevant to this paper. In \S\ref{sec:tmcsystem}, we describe the system architecture and components of HazeDose that we developed in our study. \S\ref{sec:tmcexperiment} presents the trial that we conducted to validate the performance of our system, and analyzes the trial results. In \S\ref{sec:tmcopti}, we introduce and evaluation the dosage minimization algorithm, while the paper is concluded in \S\ref{sec:tmcconclusion}.

\section{Related Work}\label{sec:related work}
Assessing the relationship between exposure level and health risks is a major research topic in both medical and engineering research area. The traditional way to estimate the relationship between air pollution exposure and health risks is to find the association between the pollution concentrations near certain locations (such as the patient's home, work, etc.) and medical adverse events (such as heart attack or emergency hospitalization) \cite{Beverland2012530, McKenzie201279, AllenMongolia, Hoek01092012, Franklincommunity, Heinrich01032013,DONS2017S94,Khafaie2017}. However, using pollution concentrations at home or work as the individual's exposure data can be erroneous, particularly for people who spend a significant fraction of the day at other places, and for people who have long commutes in traffic, during which they may inhale much of their daily pollution dose. In \cite{PARK201785}, the authors compared four different types of exposure estimates generated by using (1) individual movement data and hourly air pollution concentrations; (2) individual movement data and daily average air pollution data; (3) residential location and hourly pollution levels; and (4) residential location and daily average pollution data. They have shown that these four estimates are significantly different, which supports the argument that ignoring the spatio-temporal variability of environ- mental risk factors and human mobility may lead to misleading results in exposure assessment.

Several prior studies have included real-time location information in estimating air pollution exposure. A research group in Barcelona, Spain, designed a survey that tried to compare the exposures with different travel modes in \cite{deNazelle2012151}. They asked commuters to use different transport modes going along the same route to find out their relative inhalation dose. The inhalation rate algorithm they used was developed by other researchers, which assumed that inhalation rate ratio between different travel modes were constants. We believe that their referenced inhaled dose was neither real-time nor sufficiently accurate. They published their latest findings in \cite{JERRETT2017286}, in which they have asked participants to carry portable pollution monitors to assess the validity of a low cost personal air pollution sensor. They have shown that correlations between the personal sensors and more expensive research instruments were higher than with the government monitors, and the sensors were able to detect high and low air pollution levels in agreement with expectations. There are a few other studies that also have involved volunteers carrying portable pollution monitors. For example, the authors of \cite{Dons20113594} designed a study to find out the impact of time-activity patterns on personal exposure. They followed 16 participants, obtaining their temporal-spatial information with a PDA, and black carbon concentrations with a portable monitor. Their results showed that transportation contributed the highest black carbon concentrations. However, their study ignored the human activity levels and only estimated the pollution concentration around the participants rather than their personal inhaled dosage.

In recent years, researchers begin to utilize both real-time location and activity data to estimate exposure levels and health risks. For example, authors in \cite{Susanne} discussed how to combine individual time-activity patterns and air pollution concentrations, and gave a model to integrate the data. In \cite{6529500}, the author designed a system called ExposureSense which can combine smartphone accelerometer, external air quality data, and pluggable sensors for personal pollution exposure estimation. In these two papers, only acceleration information was considered as activity data. The research group in Spain \cite{deNazelle20092525} has also built a model for the analysis of competing risks associated with the built environment and its transformation to be more pedestrian friendly. The model used activity pattern, location, and travel mode to derive energy expenditure data, and modeled air pollution data was used to estimate inhalation dose. The results indicated a pedestrian-friendly environment would cause lower average exposure while increasing energy expenditure overall, which increased inhalation dosage.

Some research groups are using respiratory rate measurements (or estimated values) as the activity parameters. For instance, researchers in \cite{IntPanis20102263} compared vehicle exhaust air pollution exposure between car passengers and bikers. PM$_{10}$ and PM$_{2.5}$ were the pollutants considered, and the data was collected using portable optical dust monitors. Minute ventilation (VE), which was obtained by a portable cardiopulmonary indirect breath-by-breath calorimetry system, was used to calculate inhaled dose, and they concluded that inhaled particular matter (PM) was significantly higher while riding a bicycle compared to driving a car. Similarly, researchers proposed a method for monitoring and estimating people's PM2.5 inhalation by using personally physiological data and mobile PM2.5 sensor in \cite{8122716}. They retrieved people's heart rate from company's cloud platform of smart bracelet and then calculate their PM2.5 inhalation based on personal HR and minute ventilation (VE).

Some other researchers use energy expenditure as the parameter to estimate exposure levels. In \cite{00140130600708206}, the authors designed a trial to determine the level of energy expenditure and exposure to air pollution for cyclists. This study consisted of laboratory measurements and field measurements. In the laboratory part, the relationship between heart rate and pulmonary ventilation were established. In the field measurements part, heart rate was measured by heart rate monitors, while PM$_{10}$ and NO$_2$ were recorded by dust monitors. In contrast to this study, the authors of \cite{Morabia200972} assessed personal exposure to PM$_{2.5}$ and physical activity energy expenditure rate for transportation by car, train, or walking. Twenty participants who each carried an air quality monitor and a GPS receiver traveled on intended appropriate routes by car, train, and walking on three different days. Energy expenditure rates were calculated by activity metabolic equivalent (MET), speed, and body weight. These two studies, however, lacked personal inhalation dose, despite acquiring the energy expenditure data.

\begin{figure*}[!t]
  \centering
  \includegraphics[scale=0.6]{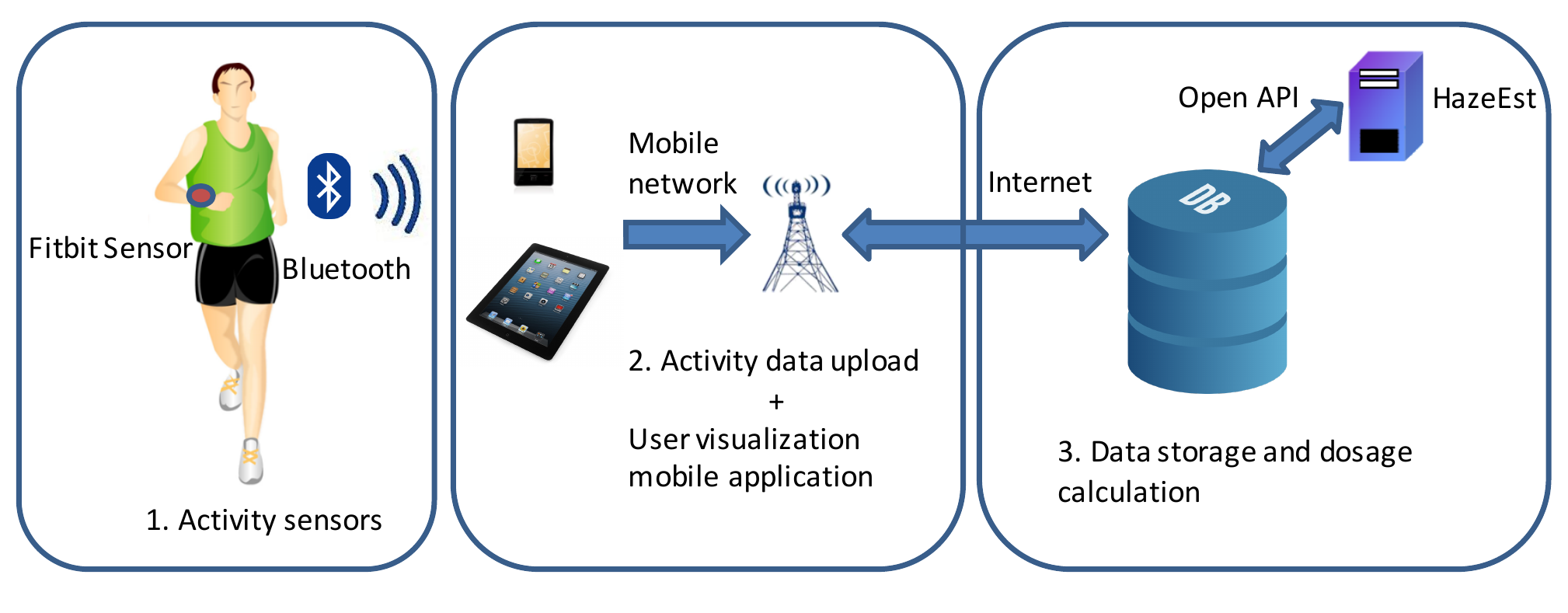}
  \vspace{-4mm}
      \caption{System architecture}
      \label{fig:tmcsys}      
\end{figure*}

In \cite{doi:10.1021/acs.est.6b05782}, the authors evaluated five methods of estimating the inhaled dose of air pollution by using a real-life data set of 122 adults who wore devices to track movement, black carbon air pollution, and physiological health markers for 3 weeks in three European cities. These five methods are: 1) methods using (physical) activity types, 2) methods based on energy expenditure, METs (metabolic equivalents of task), and oxygen consumption, 3) methods based on heart rate or 4) breathing rate, and (5) methods that combine heart and breathing rate. They concluded that there is no single best method, and the choice of a suitable method for determining the dose in future studies will depend on both the size and the objectives of the study.

As mentioned above, there are some problems with the current monitoring systems: low-resolution monitoring data from fixed stations; neither continuous nor long-term dense mobile data; lack of personal and activity data to estimate individual's exposure. All these problems lead to an insufficient accurate estimation of exposures for individuals, even to wrong medical inferences and health risk assessments.

\section{System Architecture}\label{sec:tmcsystem}
This section illustrates our HazeDose system architecture. As shown in Fig.~\ref{fig:tmcsys}, the system consists of three parts: activity sensors, data upload and user visualization mobile app, and cloud server.

\subsection{Activity Sensors}

\begin{figure}[!t]
  \centering
  \subfigure[]{
    \includegraphics[scale=0.2]{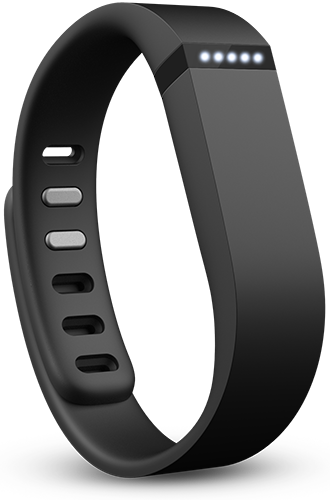}
      \label{fig:fitbit}}
        \quad
  \subfigure[]{
    \includegraphics[scale=0.07]{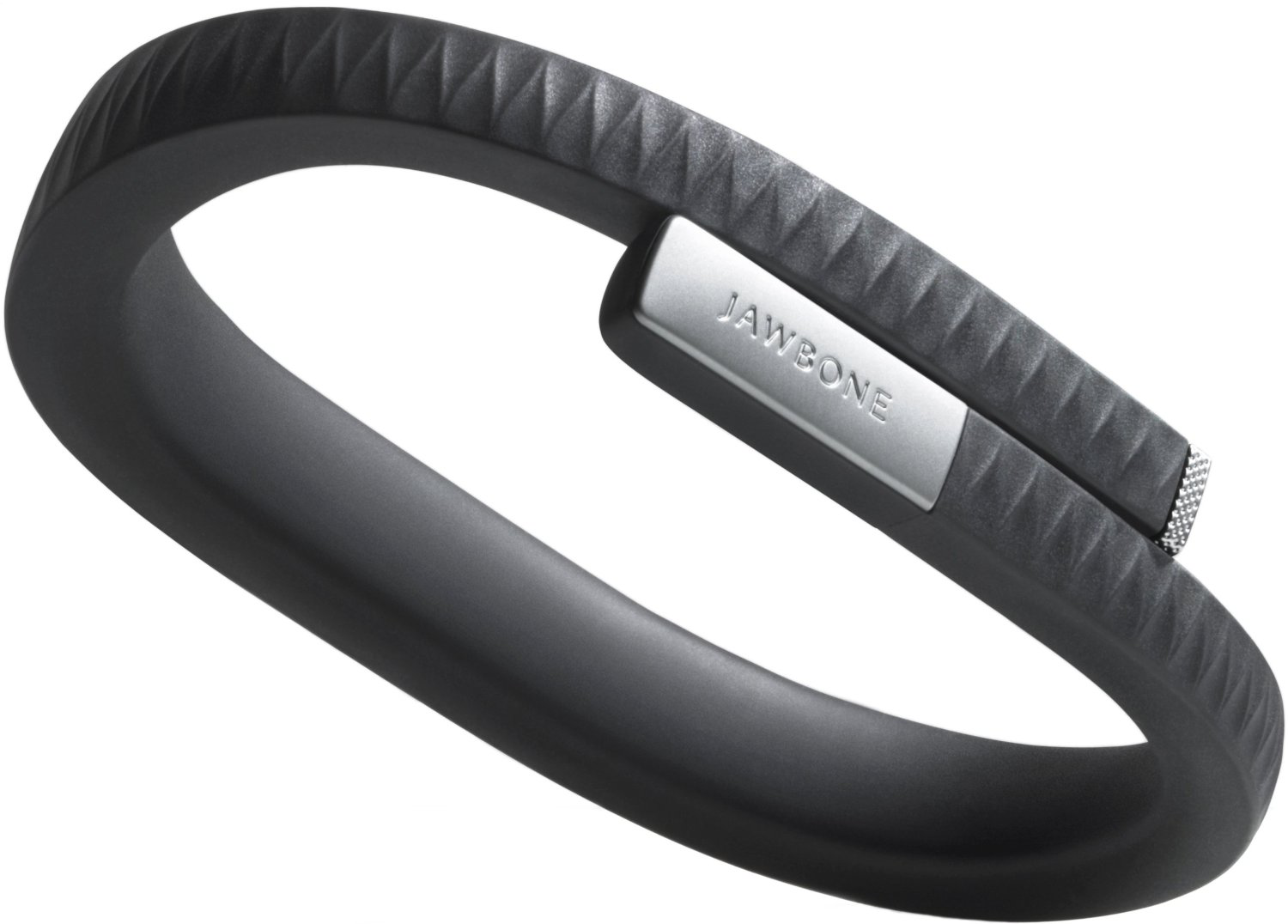}
      \label{fig:jawbone}}
        \quad
        \vspace{-3mm}
    \caption{(a) Fitbit Flex sensor, and (b) Jawbone UP sensor.}
    \vspace{-3mm}
    \label{fig:activitysensor}
\end{figure}

The chosen activity wearable sensor is the Fitbit Flex, which can acquire energy expenditure rate. We selected two activity sensors which can record energy expenditure data in the first place. One was Jawbone UP, and the other was Fitbit Flex as shown in Fig. \ref{fig:activitysensor}. Both sensors were desirable because they were built on a social platform, which encourages the user to wear the sensor continuously. While the battery life of UP was longer than that of Flex, the battery life of Flex was also over a week and thus considered acceptable. Both sensors did not provide a way to access accelerometer data directly, and they could offer information stored on their server via an open API. The single aspect that made UP unattractive to this study was that it needed to be connected to the audio jack of the iPhone to sync the data from the sensor. However, with Fitbit Flex, the users can sync the data using Bluetooth. The Jawbone UP released a new product named UP24 recently which can also support Bluetooth.

Fitbit Flex trackers use a three-axis accelerometer which can convert movements of a person into digital data to record personal motions. Fitbit company also developed a tuned algorithm to distinguish the motion patterns.

\subsection{Mobile Application}\label{mobileapps}
\subsubsection{\uline{Development Platform}}
We developed our mobile application of exposure estimates on the iOS platform. Development on Android platform requires that the application should be designed for multiple screen sizes, and it must support a spectrum of operating system versions to reach the full market. The Android version of our mobile application is possible in future work.

\subsubsection{\uline{Application Details}}
The mobile application connects to with Fitbit sensors using published API. It allows users to pair their iPhone with activity sensor over Bluetooth and get real-time data. The app will track the location data of users. All GPS data collection rely on mobile phones. Upon connecting with the selected sensors, it will periodically sample, upload and display the information as the current location on a map, the GPS information, CO concentration, and energy expenditure rate after commencement of recording. All data is recorded every minute and will be uploaded to the cloud server later to calculate personal inhaled dosage.

\begin{figure*}[!t]
  \centering
  \subfigure[]{
    \includegraphics[scale=0.3]{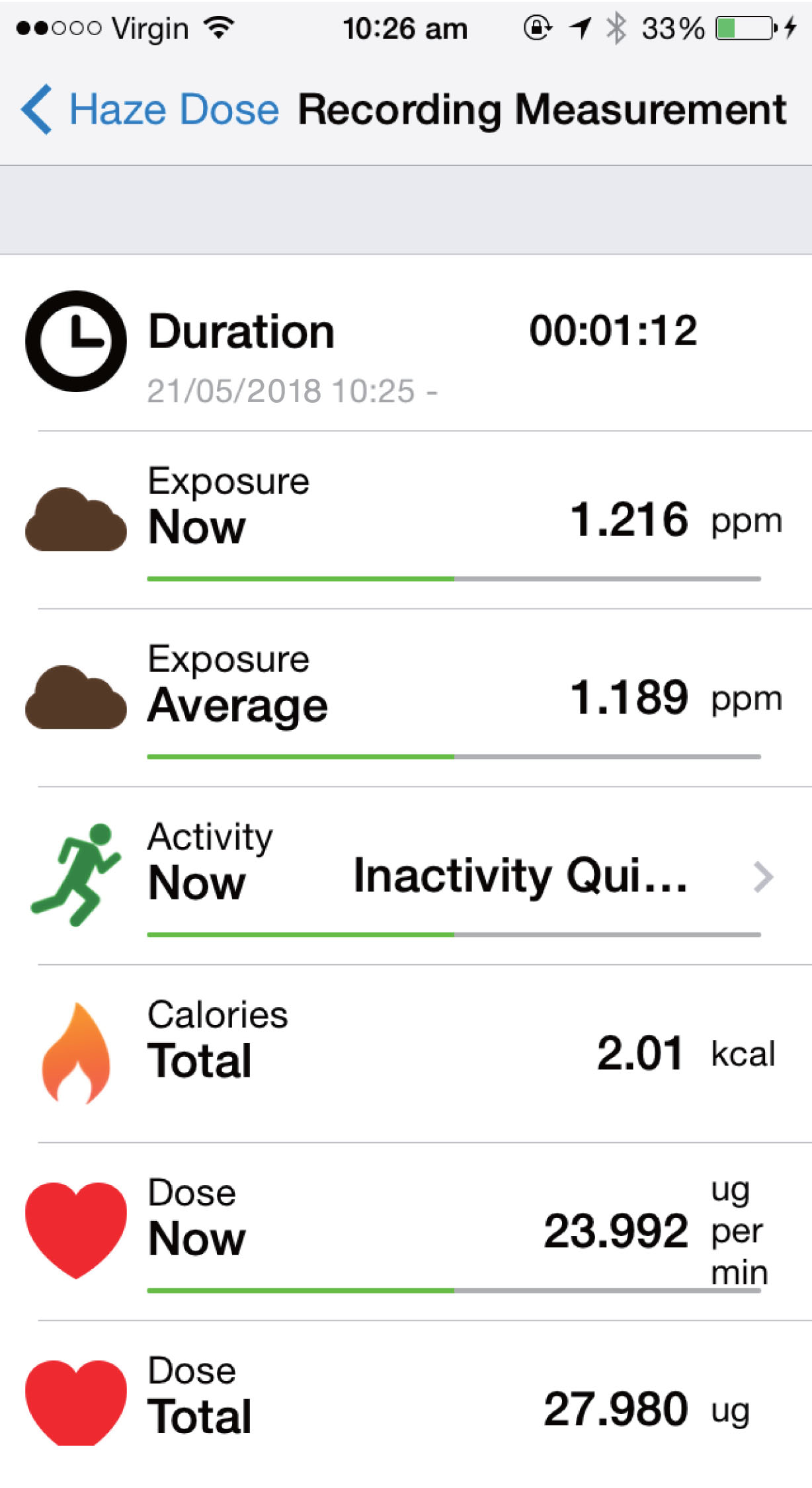}
      \label{fig:tmcsubfigure1}}
        \quad
  \subfigure[]{
    \includegraphics[scale=0.144]{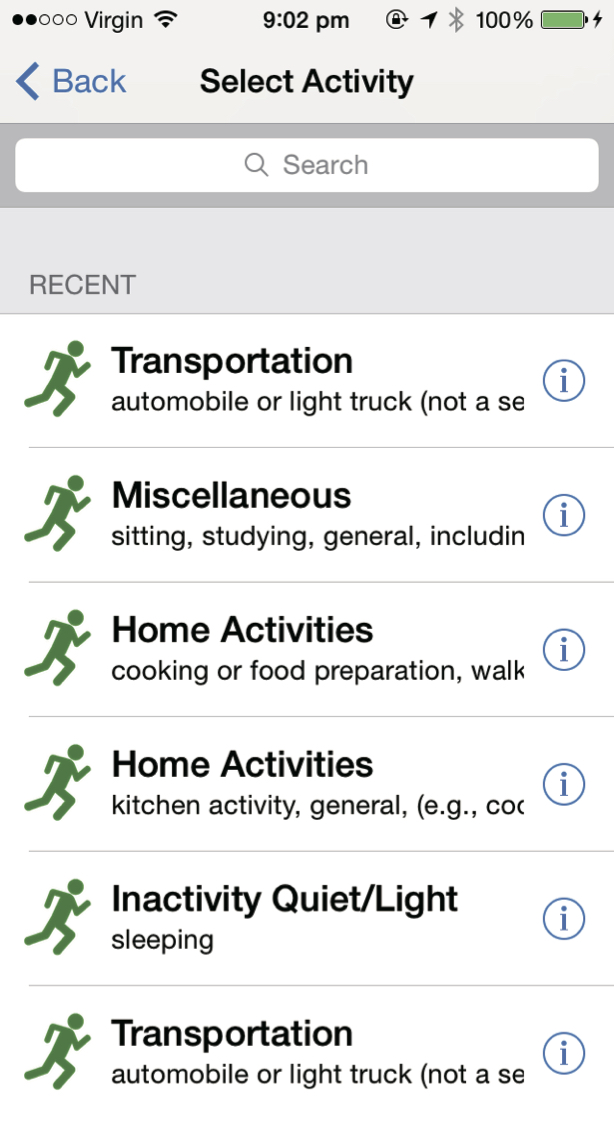}
      \label{fig:tmcsubfigure2}}
        \quad
    \subfigure[]{
    \includegraphics[scale=0.30]{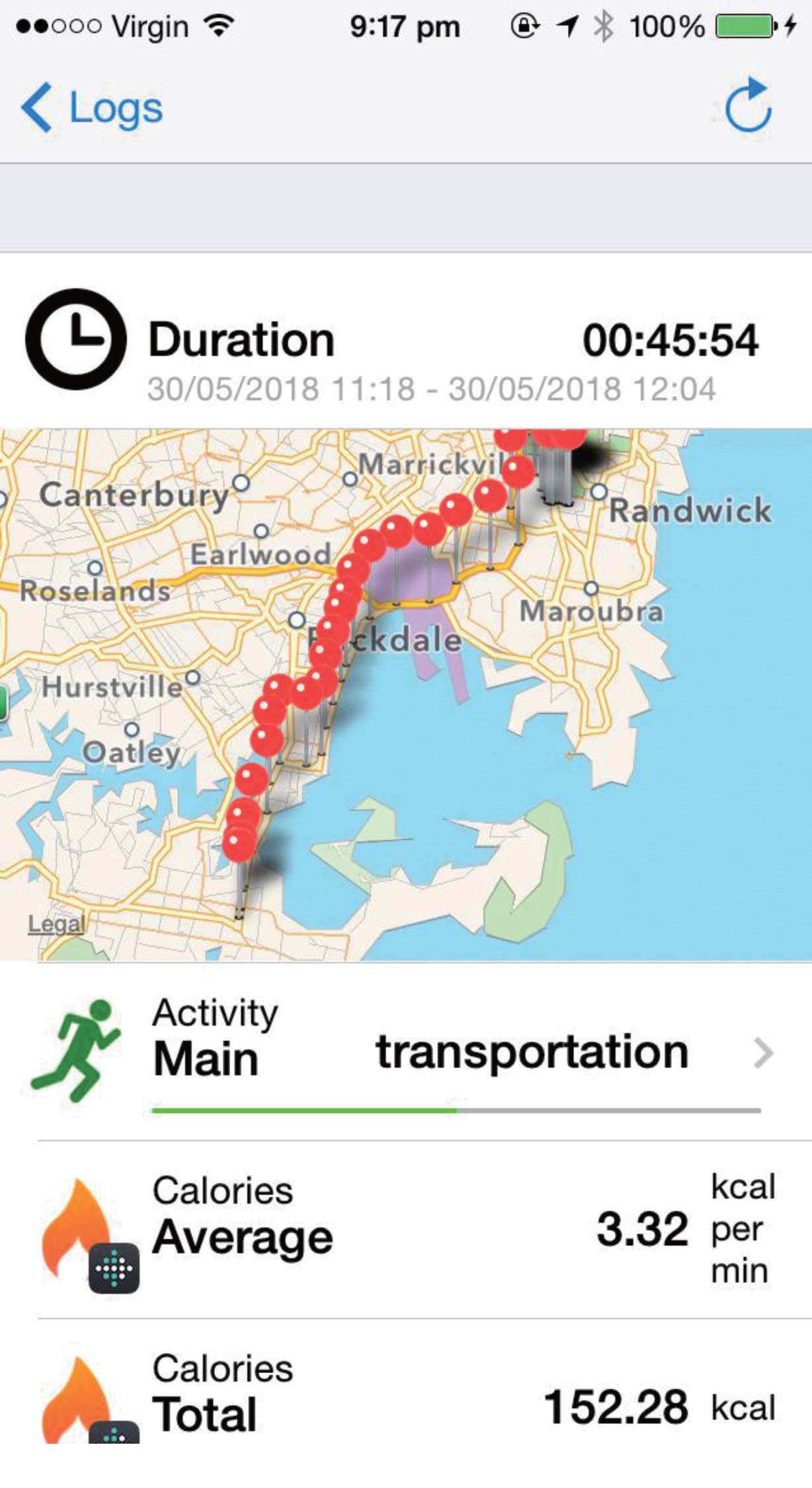}
      \label{fig:tmcsubfigure3}}
        \quad
  \subfigure[]{
    \includegraphics[scale=0.30]{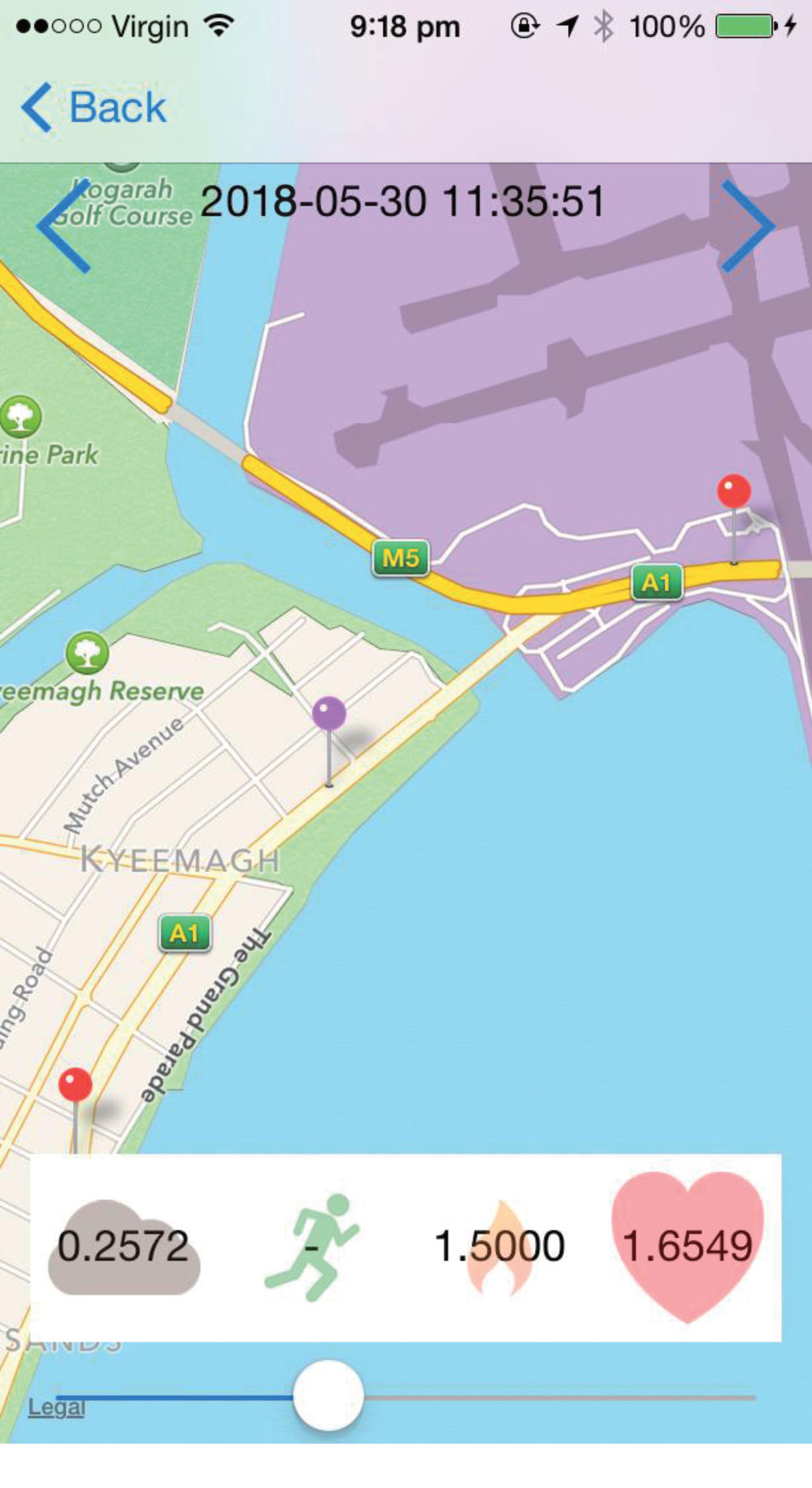}
      \label{fig:tmcsubfigure4}}
  \vspace{-3mm}
    \caption{Mobile application interface of (a) recording measurement, (b) activity mode selection, (c) log of measurement, and (d) log map}
      \vspace{-5mm}
    \label{fig:tmcfigure5}
\end{figure*}

\subsubsection{\uline{Application Demonstration}}
On first-time application launch, the first step is setup, in which the Fitbit user would be asked to log into his/her Fitbit account to synchronize the age, body mass, and gender from the Fitbit server. Alternatively, if the user does not use Fitbit, these details can be obtained in the setup process manually. It must be pointed out that an accurate value for personal information like body mass is not guaranteed, as it depends on on whether users diligently update their details on the application or via Fitbit. In the case of a negligent user, if the error between the recorded body mass and actual body mass is 5kg, it will result in an error of up to 10\% in the final calculated dose. Such a large error is undesirable and cannot be corrected for within our application.

When the setup is complete, the user can just tap Start Measurement icon to start recording. Fig. \ref{fig:tmcsubfigure1} shows the app interface while “Recording Measurement” is ongoing. It displays the statistics of user data (e.g. real-time and average exposure, ...) from the commencement of
recording. Meanwhile, users without a Fitbit can select activity mode to get estimated energy expenditure data in the recording page. Fig. \ref{fig:tmcsubfigure2} illustrates the selection of Activity mode on the app interface. Note that users need to enable the GPS service on their mobile phone to provide the app with current location. The mobile application acquires the air pollution data from our server along with the location and time stamp. Users with a Fitbit cannot get real-time dosage estimation in the measurement page as the Fitbit Flex needs to sync with its app before energy expenditure rate data can be retrieved from the Fitbit servers. After a measurement recording is finished, the user can view all the recorded data in the log menu, as shown in Fig. \ref{fig:tmcsubfigure3}. Each log can present information including recorded duration, route tracking, calories burned, and total dosage. A non-Fitbit-user can just get personal dosage data from the log interface, while a Fitbit user has to synchronize with Fitbit server using his/her Fitbit mobile application to get past 7 days energy expenditure data by Kcal per minute. The map in the log view can be tapped for a
transition to Measurement Map view as shown in Fig. \ref{fig:tmcsubfigure4}. In Measurement Map view, the user is provided with a large map, which, at first, shows the user location recorded every minute – red pins on the map represent user locations. If a pin is tapped, it gets selected and turns purple. A popup box then appears that displays air pollution concentration level, the activity level in Metabolic Equivalent (MET), calories burned, and dosage of that minute. All the measurements data, including personal information, activity modes, calories burned, locations, and dosages, can be sent to the user's email address as attachments (csv format), if the user is interested in the data and would like to explore the data in details.

\subsection{Cloud Server}
This is the central repository, to which our users upload energy expenditure data into a MySQL database. For privacy protection, no personal energy expenditure data is shared with other users. We compute the personal inhaled dosage using an algorithm (explained next) with inputs from the HazeEst system (via open API) and activity data. HazeEst system is a dense air pollution estimation system which uses historical data from both (sparse) government monitoring sites and (dense) wireless sensor network and Support Vector regression (SVR) model to learn air pollution profile at fine spatial granularity, and thereafter estimates the air pollution surface for any day/time in metropolitan Sydney. The computed dosage value is fed back to the user mobile app to visualize the personal exposure information. 

\textbf{Dosage Calculation algorithms:}
We used an algorithm discussed in \cite{Johnson} to convert energy expenditure rate to inhalation dose. We summarize the algorithm in Table \ref{table:Dosagemodel}. The main idea of the algorithm is to estimate Energy Expenditure (EE) rate based on personal and activity data, and convert EE rate to breathing rate, then calculate personal dosage using both breathing rate and air pollution concentration data. The whole algorithm comprises of two parts, which enables both activity sensor users and non-activity-sensor users to benefit from our system. Energy expenditure estimation part is not necessary for activity sensor user as energy expenditure data can be acquired from activity sensor server.

\section{Dosage Estimation Trial}\label{sec:tmcexperiment}
To evaluate the performance of our system, we designed a trial and involved several participants in this study between 10/09/2016 and 16/09/2016. 

% \begin{table*}[!t] \renewcommand{\arraystretch}{1}
%   \caption{Estimate inhalation dosage algorithm}
  \begin{longtable}{ p{15.5cm} }
  \caption{Estimate inhalation dosage algorithm}\\
  \hline
  \vspace{0.01cm}
  1.Estimate Energy Expenditure\\
  1.1. Calculate Resting Metabolic Rate (RMR): \\
  \begin{equation}\label{equation1}
RMR = (0.166)*[a+b*(BM)+e],
\end{equation} \\
  where a and b are constant appropriate regression parameters and determined by age and gender, and BM is body mass (kg) while e is a randomly selected value from a normal distribution with mean equal to zero. 0.166 converts the unit of RMR from MJ day$^{-1}$ to Kcal min$^{-1}$.\\
  1.2. Calculate Energy Expenditure (EE):\\
  \begin{equation}\label{equation2}
EE = MET * RMR,
\end{equation} \\
    in which MET is Metabolic Equivalent. The MET would be different based on various physical activities and listed in \cite{ainsworth2000compendium}. EE is also expressed in Kcal min$^{-1}$. \\
  2.Calculate Inhalation Dosage (ID):\\
  2.1. Calculate Oxygen Uptake Rate (VO$_2$):\\ 
  \begin{equation}\label{equation3}
VO_2=ECF*EE,
\end{equation} \\
  in which ECF represents Energy Conversion Factor defined as the volume of oxygen
required to produce one kilocalorie of energy and has the unit of liters oxygen Kcal$^{-1}$. ECF is unique to diverse people and a random variable of uniform distribution between 0.20 and 0.21. 0.205 is applied in our application. VO$_2$ has the unit of liters oxygen min$^{-1}$.\\
  2.2. Calculate Ventilation Rate (VR):\\
  \begin{equation}\label{equation4}
VR=BM*\mathrm{e}^{c+d*\ln\frac{VO_2}{BM}},
\end{equation} \\
  where VR is measured in liter min$^{-1}$ and constants c and d are also determined by age and gender.\\
  2.3. Calculate Inhalation Dosage (ID):\\
  \begin{equation}\label{equation5}
ID=VR*PC,
\end{equation} \\
in which PC represents air pollution concentrations in $\mu$g liter$^{-1}$.\\
  \hline
  \label{table:Dosagemodel}
  \end{longtable}
%     \label{table:Dosagemodel}
% \end{table*}

\begin{table*}[!t] \renewcommand{\arraystretch}{1}
  \caption{Participant general information}
  \begin{tabular}{ p{2cm} p{3cm} p{2.8cm} p{3cm} p{3cm}}
  \hline
  Gender & Body mass (kg) & Age & Stature(cm) & Identity\\
    \hline
  Male &  78.6 &  33 & 180 & Walker 1\\
  Male &  67.3 &  25 & 179 & Walker 2\\
  Female &  46.7 &  36 & 160 & Biker 1\\
  Male &  81.1 &  44 & 184 & Biker 2\\
  Male &  64.0 &  43 & 182 & Driver 1\\
  Female &  51.9 &  27 & 167 & Driver 2\\
  \hline
  \end{tabular}
    \label{table:participant}
\end{table*}

\begin{figure*}[!t]
  \centering
  \includegraphics[scale=0.25]{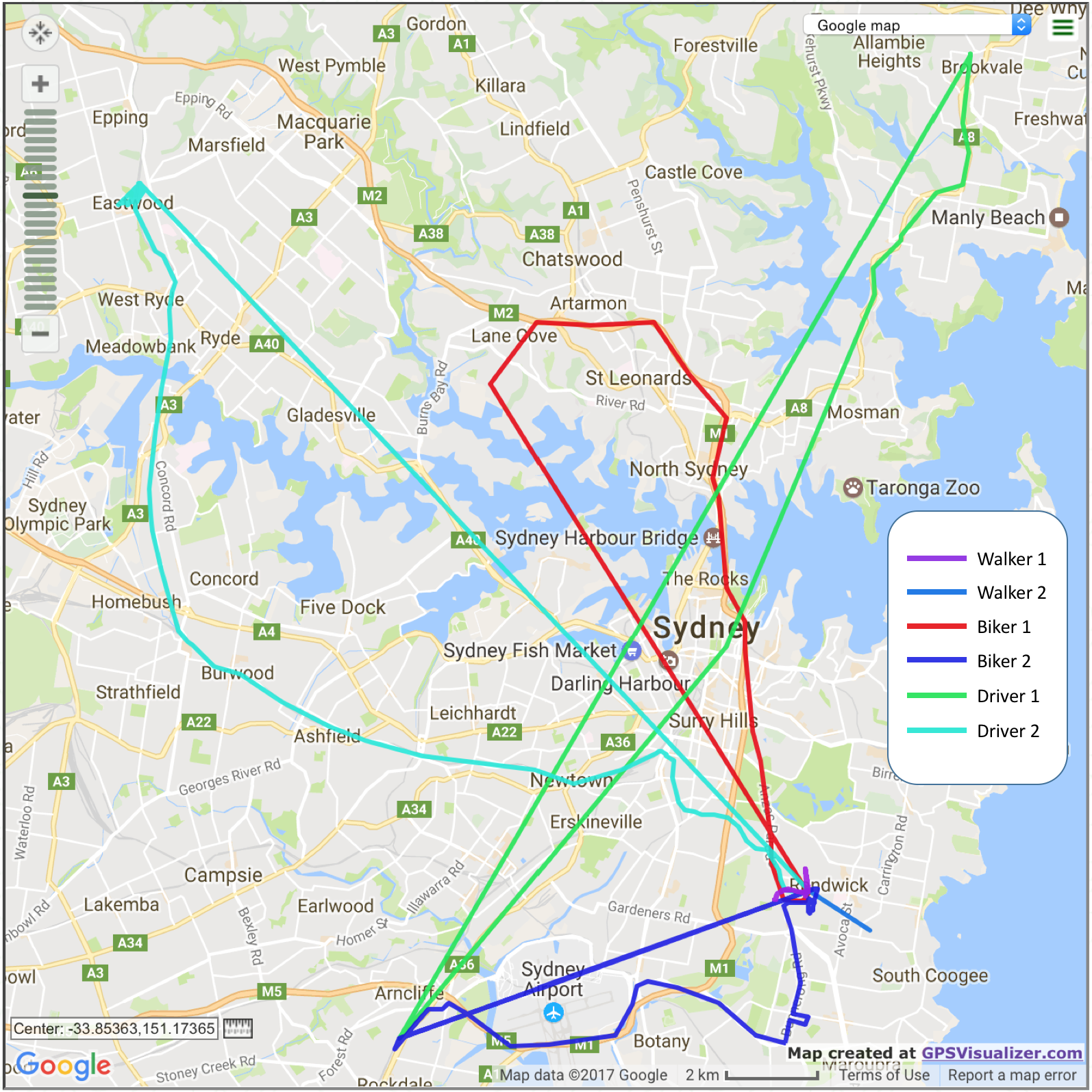}
      \caption{Commuting routes taken by different participants.}
      \label{fig:commuteroutes}
\end{figure*}

\subsection{Participants}
Six volunteers who work or study in UNSW Australia participated in this study. They were put into three groups based on their commuting patterns (walking, cycling, or driving). The walking participants lived more than a 15-minute walk away from the university, while the commuting time of cycling and driving participants can be up to 1.1 hours one-way. Table \ref{table:participant} gives the general information of these participants.

\subsection{Trial Procedures}
While participants followed their normal schedule, they were asked to wear the Fitbit sensor during the trial week and keep recording the location and energy expenditure data using our mobile application. The data was recorded every minute as described in section \ref{mobileapps}. We also asked all the participants to start recording at 8:00 am and end recording at 11:59 pm every day, and the recorded data was uploaded to our server automatically when the user stopped recording. Further, the participant had to change activity modes manually to get constant EE data for different activity types. Fig. \ref{fig:commuteroutes} shows the map of commuting routes followed by different participants, and different color indicates the different person. The routes which cover an urban area of approximately 1000km$^2$ contains motorway, tunnel, footway, campus, and residential areas.

 \begin{table*}[!t] \renewcommand{\arraystretch}{1.1}
 \newcommand{\tabincell}[2]{\begin{tabular}{@{}#1@{}}#2\end{tabular}}
  \caption{Samples of collected data}
   \centering
  \begin{tabular}{p{3cm} p{4cm} p{3.8cm} p{3.5cm}}
  \hline
      Time & Activity modes & Latitude & Longitude \\
    \hline
     15:43:00 & Work & -33.91751464 & 151.23194040 \\
  \hline
  \end{tabular}
  \begin{tabular}{p{7.5cm} p{7.5cm}}
     Manual EE (Kcal per min) & Fibit EE (Kcal per min)  \\
    \hline
      1.43 & 5.51 \\
  \hline
  \end{tabular}
    \begin{tabular}{p{7.5cm} p{7.5cm}}
     Manual dosage ($\mu$g per min) & Fitbit dosage ($\mu$g per min)\\
    \hline
     0.23 & 1.14\\
  \hline
  \end{tabular}
    \label{table:Samples}
\end{table*}

\begin{table*}[!t] \renewcommand{\arraystretch}{1.1}
\newcommand{\tabincell}[2]{\begin{tabular}{@{}#1@{}}#2\end{tabular}}
  \caption{Trial result attributes}
   \centering
  \begin{tabular}{p{2cm} | p{3.6cm} p{3.6cm}| p{5cm}}
  \hline
      Participants &  \multicolumn{2}{|c|}{EE (Kcal per min) (Average (Min - Max))}   &  CO concentrations (ppm)\\   
    \hline
      & Manual  & Fibit &\\
    \hline  
     Walker1 &   1.96(1.52 - 5.17) &  1.65 (1.10 - 9.26)&   0.57 (0.16 - 7.50)\\
     \hline
     Walker2 &   1.71(1.47 - 5.01) & 1.52 (1.17 - 7.75)&   0.52 (0.13 - 5.58)\\
     \hline
     Biker1 &   2.43 (1.51 - 8.98) & 2.02 (1.10 - 10.27)&   0.71 (0.11 - 12.96)\\
     \hline
     Biker2 &   2.87 (1.69 - 9.55) & 2.05 (1.25 - 14.85)&   0.76 (0.11 - 14.93)\\
     \hline
     Driver1 &   1.61(1.43 - 2.19) & 1.03 (0.78 - 7.57)&   0.88 (0.18 - 13.08)\\
     \hline
     Driver2 &   1.53(1.36 - 2.08) & 1.49 (0.95 - 8.00)&   0.81 (0.12 - 17.89)\\
  \hline
  \end{tabular}
    \label{table:attributs}
\end{table*}

\subsection{Results and Discussion}\label{sec:tmcresults}
\subsubsection{\uline{Collected data}}

All the participants strictly followed our guidance and each of them successfully uploaded seven days of data. Table \ref{table:Samples} gives a sample of what we have collected during the seven-day period. Manual EE is calculated using MET data as mentioned in section \ref{mobileapps}, while Fitbit EE is downloaded from the Fitbit server directly.

Table \ref{table:attributs} shows the result attributes of the seven-day trial study, including the measured and manual selected energy expenditure of the participants, and CO pollution concentrations. It is seen that ambient CO concentration data from our HazeEst system shows a significant variation (range from 0.11ppm to 17.89ppm), with pollution peaking along the traffic jam section which reaches to a high level (walking: 7.50; bicycling: 14.93; driving: 17.89 ppm). The highest mean CO concentrations exposure by the participants is 0.88 ppm, and we observe that CO concentrations are significantly higher for driving and cycling compared to walking, since walkers all live near campus which is far away from heavily congested roads or highways. This is consistent with our expectation that on-road drivers will experience higher pollution concentrations that off-road bicyclists and joggers. Meanwhile, we can see that the mean energy expenditure data of bikers is greater than walkers and drivers. The peak EE of bikers can be up to 14.85 Kcal per minute, while it is 9.26 and 8.00 for walkers and drivers respectively. Similarly, the bikers have almost double average EE values compared to drivers.

\begin{figure*}[!t]
  \centering
  \includegraphics[scale=0.7]{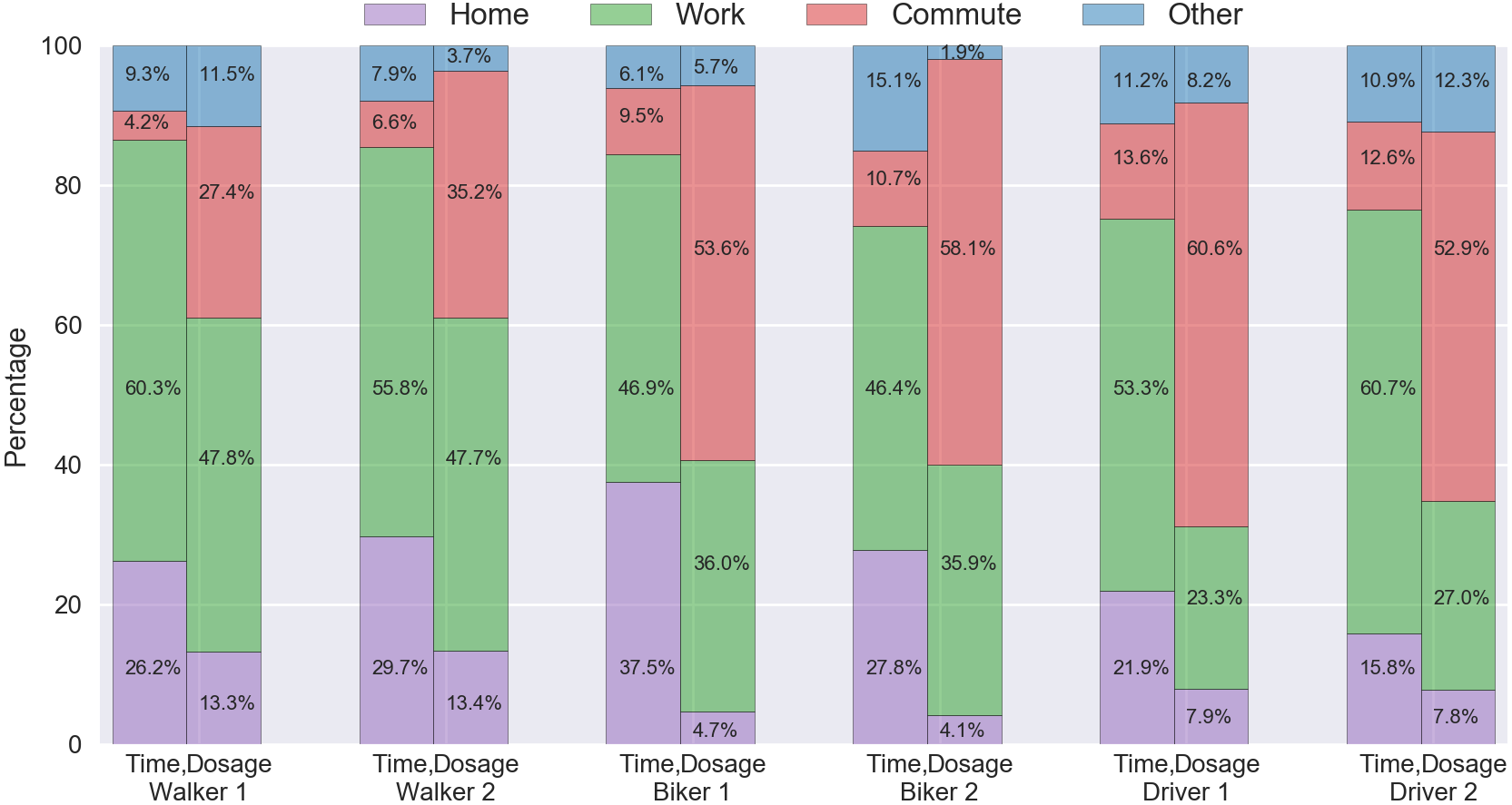}
      \caption{Time spent and CO dosage proportion in different locations of six participants.}
      \label{fig:bar}
\end{figure*}

\subsubsection{\uline{Time-Dosage profile}}

\begin{figure*}[!t]
  \centering
  \subfigure[]{
    \includegraphics[scale=0.35]{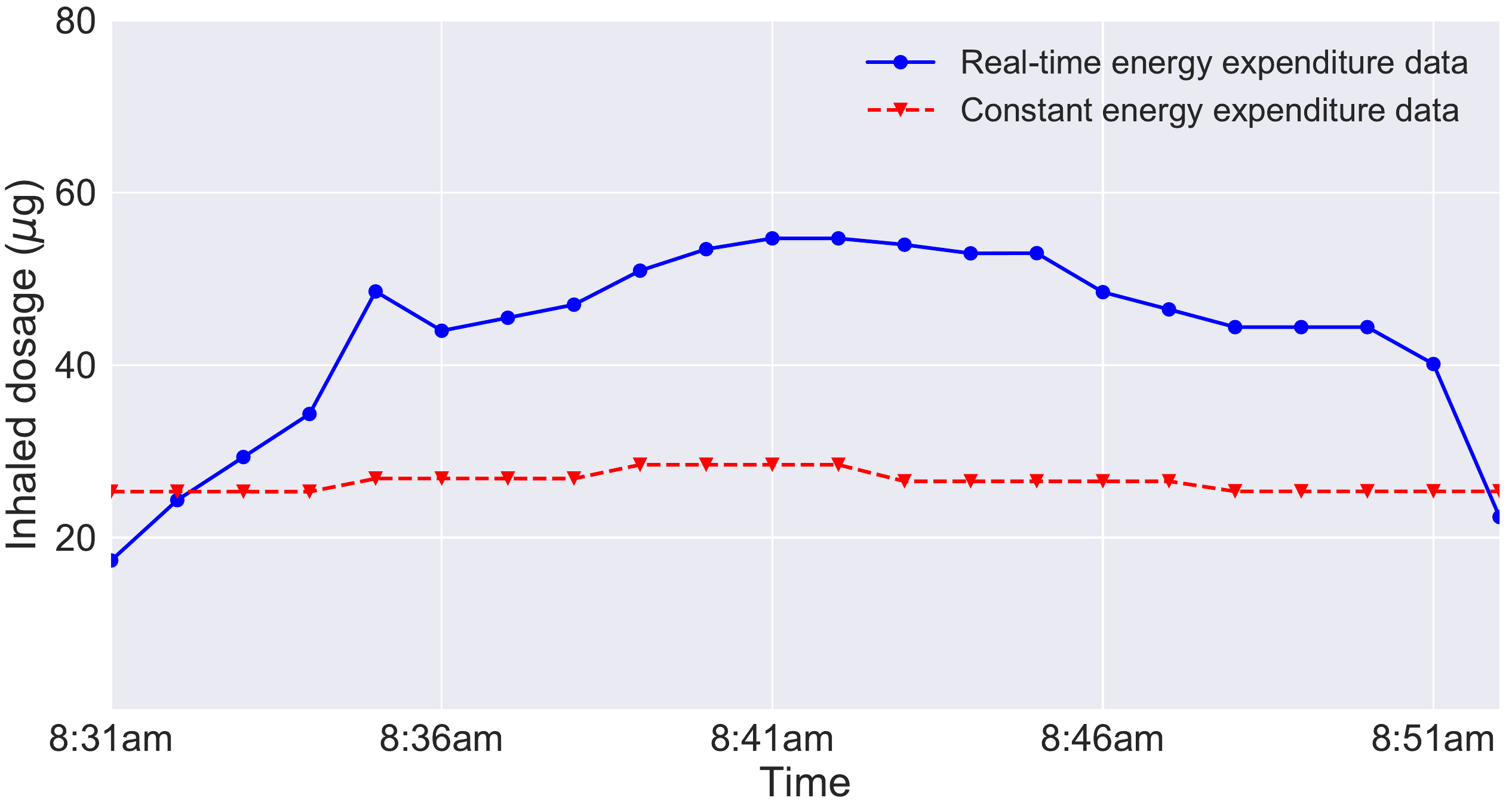}
      \label{fig:walkerdosage}}
        \quad
    \subfigure[]{
    \includegraphics[scale=0.35]{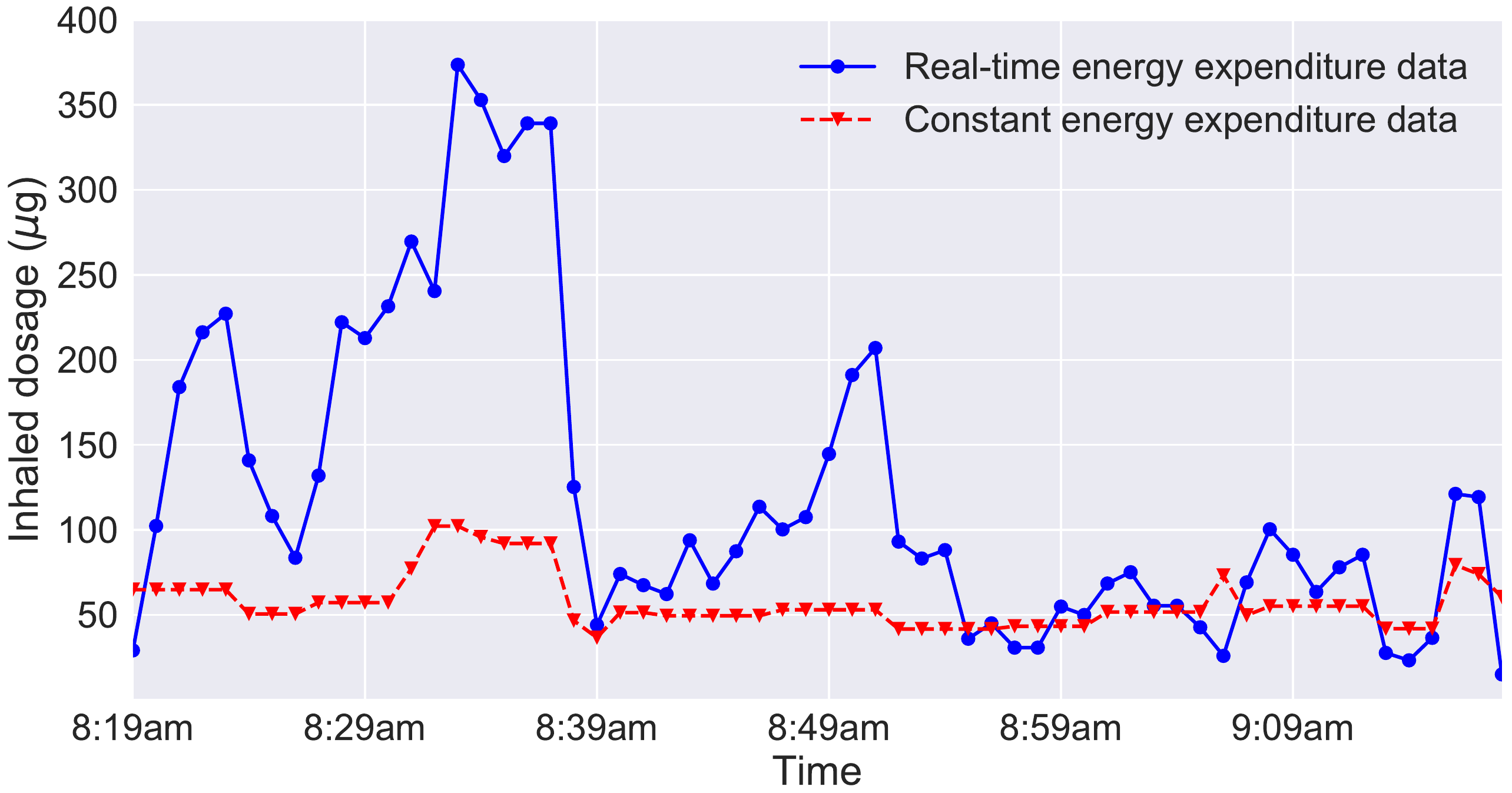}
      \label{fig:bikerdosage}}
        \quad 
    \subfigure[]{   
    \includegraphics[scale=0.35]{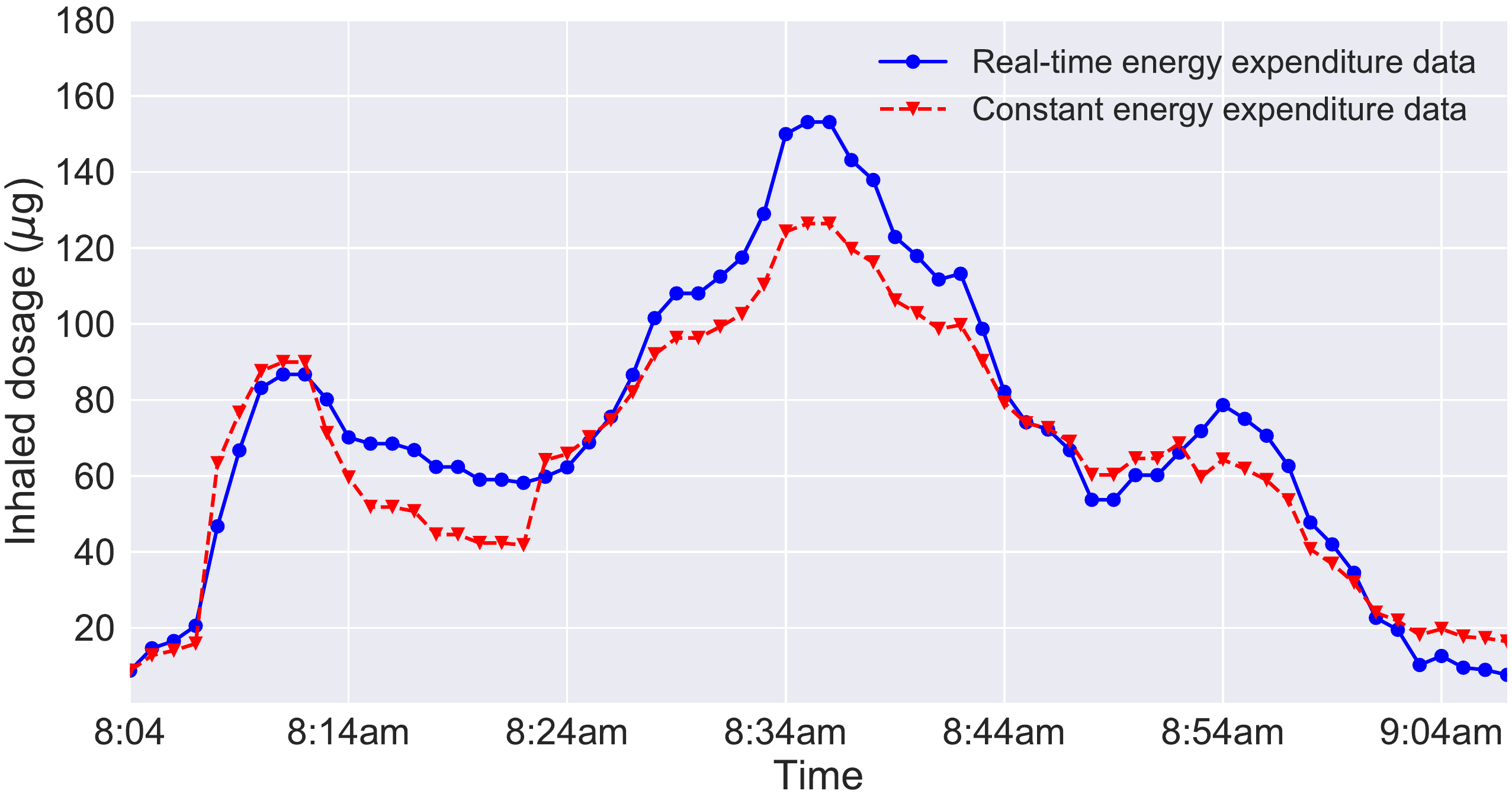}
      \label{fig:driverdosage}}
        \quad    
    \caption{Carbon monoxide inhaled by (a) Walker 1, (b) Biker 1, and (c) Driver 1 during commuting.}
    \label{fig:doagecompare}
\end{figure*}

Since air pollution data from our HazeEst system is available from 8:00 am to 12:00 am, we only profile time spent and dosage proportion in different locations during that period. As we can see from Fig. \ref{fig:bar}, all the participants spent on average almost half of their time at work (shown by green bars), and inhaled 23.3\% - 47.8\% of their daily dosage (not including eight hours of sleeping time at home). Commuting time, by contrast, takes 4.2\% to 13.6\% of their time, but seizes 27.4\% - 60.6\% of the daily inhaled dose (shown by red bars). For example, Biker 2 spent on average more than one forth and almost half of his time at home and workplace respectively, while it only takes 10.7\% of his time in commuting. However, the inhaled dosage occupies more than half of his average daily dosage, while he exposes 44.0\% at home and work.

This profile indicates that commuting air pollution exposure is non-negligible. On the contrary, it takes a significant proportion of people's daily exposure. It proves that only using the air pollution concentrations near certain locations (such as patient's home location, work location, etc.) to estimate the relationship between the air pollution exposure and medical adverse events (such as heart attack or emergency hospitalization) may lead to biased medical inferences.

\subsubsection{\uline{Activity-Dosage profile}}

To better understand how different activities impact personal exposure, we plot commuting inhaled dosage estimations of three participants (Walker 1, Biker 1 and Driver 1) in one morning during the trial as shown in Fig. \ref{fig:doagecompare}. For one participant, we compare the dosage estimations using real-time EE data from Fitbit sensors and constant EE data using the MET model. 

First, we see that with constant EE data, inhaled dose estimations are at around 25 $\mu$g per minute for Walker 1 along the whole commuting route. The reason is that Walker 1 lives near campus and the pollution concentrations do not change much from his home to the university, and leads to a flat and low inhaled dose. However, when we take real-time EE data into consideration, the estimations increase up to 55 $\mu$g per minute. Similarly, if we only use constant EE data, the dosage of Biker 1 stays lower than 100 $\mu$g per minute. However, the dosage significantly increases and peaks at 375 $\mu$g per minute when we use real-time EE data. The situation is sightly different for Driver 1. Since the energy expenditure rate would be fairly consistent during driving, while air pollution concentrations change significantly along the road/highway, the inhaled dosage estimations of using constant and real-time EE data for Driver 1 are almost the same.

Another observation is that driving incurs highest inhaled dose (130 $\mu$g min$^{-1}$), more than bicycling (100$\mu$g min$^{-1}$) or walking (25$\mu$g min$^{-1}$) when constant EE data is used. However, while taking the real-time EE into consideration, the situation reverses -- the Biker's inhaled dose per minute is higher than the Driver. This illustrates that cycling leads to higher pollution intake dosage due to increased breathing rate, compared to driving. Interestingly, bicycling turns out to be even worse than walking, probably because it happens closer to the high-traffic roads than walking while having a high energy expenditure rate.

It can be seen that our HazeDose system can generate more accurate inhaled dose estimations using real-time energy expenditure rate from the Fitbit sensor compared to using constant EE data from the MET model. Nevertheless, dosage estimations with and without a Fitbit correlate well, although they are distinct during walking and cycling as a result of activity MET estimation deviation.

\section{Dosage Minimization}\label{sec:tmcopti}
We have shown that our HazeDose system can give individuals their personal air pollution inhaled dosage estimations. To help them to manage and reduce their exposure, in this section, we propose a dosage minimization algorithm to minimize personal daily air pollution inhaled dosage by adjusting their daily schedule and commute routes.

\subsection{Minimization Model}\label{sec:tmcoptimodel}
In section \ref{sec:tmcresults}, we can see that commuting contributes to a large proportion of personal daily inhaled dosage, especially for bikers and drivers. Therefore, the dosage minimization problem is to minimize the inhaled dosage during commuting while completely satisfying the working and studying requirements. The objective function is expressed as:

\begin{equation}
\begin{split}
min\ Dose_{commuting}  = \sum_{l \in R_1}^{}\sum_{t_1 = \omega}^{\omega + h_1} D_{l_1}(t_1)\\
+ \sum_{l \in R_2}^{}\sum_{t_2 = \omega+h_1 + \beta}^{\omega+h_1 +\beta + h_2} D_{l_2}(t_2), \\ 
\label{equation:tmc1}
\end{split}
\end{equation}
where
\begin{equation}
D_{l}(t)  =  c_{l}(t)*b_{l}(t). \\ 
\label{equation:tmc2}
\end{equation}
Here, $R$ is the set of locations from home to work ($R_1$), and from work back home ($R_2$). $\omega$ and $\beta$ are the home leaving time and working hours respectively, and $h$ denotes the commute duration. $c_{l}(t)$ is the Carbon Monoxide concentration ($\mu$g/L), and $b_{l}(t)$ is the ventilation rate ($L/min$) at a particular time $t$ and a particular location $l$. Therefore, $D_{l}(t)$ is the inhaled dosage value at location $l$ and time $t$. We assume that if the individual follows the same route using one specified commuting pattern, the ventilation rate will be the same no matter when does the individual start to commute.
The objective subjects to the following constraints:

\begin{itemize}
\item $\omega$ $\in$ [$\omega_{min}$, $\omega_{max}$], where $\omega_{min}$ and $\omega_{max}$ indicates the earliest and latest home leaving time.
\item $\beta$ $\in$ [$\beta_{min}$, $\beta_{max}$] , where $\beta_{min}$ and $\beta_{max}$ indicates the minimum required working hours, and maximum hours staying in the work place respectively.
\end{itemize}

\subsection{Commuting Routes}\label{sec:tmcroutes}
As we can see from Equation \ref{equation:tmc1}, personal dosage at a certain location, home leaving time, working hours, and commute route and duration are needed to get personal commuting air pollution dosage. Personal dosage can be acquired from our HazeDose system as mentioned above, and home leaving time, working hours, commute route and duration are the tunable parameters which help to minimize dosage. Unlike leaving home time and working hours which can be decided by human, alternative commute routes and duration are often determined by the routing application programming interfaces (APIs), such as Google Maps API \cite{Boulos2005}, which usually offers two to three different routes based on the origin and destination pairs. To decide the alternative commute routes in this research, we implemented a route detection program based on the Google Maps API. 

The program is written in Python, and contains two parts to get all the feasible routes and the coordinates along an entire route. The first part is using Google Maps API, origin and destination pair, and leaving time to get the route profile. The route profile contains multiple consecutive route segments, duration, and distance features for each route segment. The second part converts these segments into real geographic coordinates. The inputs to this program include origin, destination, travel mode, and departure time. The outputs are a series of coordinates for all the possible routes. To reduce the querying time, we save all the route profile and the corresponding coordinates into our database. When the responding route profile can be found in the database, we fetch the coordinates from the database directly instead of starting another query.

\subsection{Solution Algorithm}

To get the minimum $Dose_{commuting}$, we use (a) a brute-force algorithm to enumerate all the possible commute patterns with fine time granularity to get the global optimal value, and (b) a heuristic that operates at coarse time granularity to reduce algorithm execution time at the expense of being sub-optimal.

In the brute-force algorithm, first, we assign two initial values, which are the earliest starting commuting time $\omega_{min}$ and minimum working hours $\beta_{min}$. Then we use the route detection program to get all the feasible commuting duration $h_i$ and routes $R_i$ based on the starting commuting time for both ways (home to work, and back home). After that, we compute $Dose_{commuting}$ for each route using equation \ref{equation:tmc1}. When we get these dosage values, we save them along with $\omega_{min}$ and $\beta_{min}$ in a table $T$. Next, we add one minute to the minimum working hours and get $\beta_2$. If $\beta_2$ is smaller or equal to the maximum hours staying in the workplace, we re-calculate the dosage value and save it to the existing table along with $\omega_x$ and $\beta_x$. Until $\beta_x$ reaches the maximum staying hours in the workplace, we add one minute to the earliest starting commuting time $\omega_{min}$ and repeat the whole process. The minimum dosage value in the table is the minimized dosage value. 

Solution algorithm for the heuristic algorithm is as same as for the brute-force algorithm, except we use $t$ minutes (($\omega_{max} - \omega_{min}$) $>=$ $t$ $>$ $1$) as the time granularity instead of one. The time granularity is the time interval between current and next commute pattern, and it affects both algorithm execution time and minimization results. Selecting one certain time granularity to balance the execution time and optimal result will be discussed in Section \ref{sec:tmcheuristic} below.

The overview of solution algorithm is shown is Table \ref{table:tmcsolution}.

\begin{table*}[!t]
\centering
  \caption{Solution algorithm}
  \begin{tabular}{ |p{15.5cm} |}
  \hline
 $Input$: commuting duration $h_i$; \\
  \ \ \ \ \ \ \ \ \ \ commuting routes $R_i$(per minute coordinate along the route);\\
 \ \ \ \ \ \ \ \ \ \ starting to commute time window [$\omega_{min}$, $\omega_{max}$];\\
 \ \ \ \ \ \ \ \ \ \ working hours time window [$\beta_{min}$, $\beta_{max}$];\\
 \ \ \ \ \ \ \ \ \ \ time granularity $t$.\\
 $Output$: minimized dosage value $Dose_{commuting}$; \\
  \ \ \ \ \ \ \ \ \ \ starting to commute time $\omega_x$; \\
   \ \ \ \ \ \ \ \ \ \ working hours $\beta_x$;\\
   \ \ \ \ \ \ \ \ \ \ Two Routes (home to work, work back home) $R_{1x}$, $R_{2x}$.\\
   \hline
$Step 1$: Set two initial values -- earliest starting commuting time $\omega_{min}$ and minimum working hours $\beta_{min}$. \\
$Step 2$: Get route options from the route detection program for both ways -- $h_i$ and $R_i$.\\
$Step 3$: Calculate $Dose_{commuting}$ and save it alone with $\omega$ and $\beta$ in table $T$.\\
$Step 4$: Add $t$ minutes to $\beta$ and get $\beta_x$, x $\in$ \{2,3,4,....\}. If $\beta_x$ $\leq$ $\beta_{max}$, go to step 2.\\
$Step 5$: If $\beta_x$ $>$ $\beta_{max}$, add $t$ minutes to $\omega$ and get $\omega_x$, , x $\in$ \{2,3,4,....\}.\\ 
$Step 6$: If $\omega_x$ $\leq$ $\omega_{max}$, go to step 2.\\ 
$Step 7$: If $\omega_x$ $>$ $\omega_{max}$, stop. Find the minimum dosage value in table $T$.\\ 
  \hline
  \end{tabular}
    \label{table:tmcsolution}
\end{table*}

\subsection{Evaluation of the Brute-force Solution}
We designed two different sets of scenarios to evaluate the performance of the brute-force algorithm. First, to make the scenario simpler, we assume the commuting duration $h$ and routes $R$ are known for both ways, and the minimization problem becomes a time schedule arranging problem. Second, we include alternative route option in the scenarios, and try to minimize the dosage with joint schedule and routes. All the scenario data is based on the trial data and three different commute patterns from the previous section.

In both sets of scenarios, we assume that even if we change the commuting period, the rest of the activities will not change (such as working, walking on the campus, etc.), however, the period of the rest activities (e.g. start to work time) will change along with the commute time. For example, if the time of arriving at university moves from 8:30 am to 9:00 am, the starting to work time will also move from 8:31 am to 9:01 am. Moreover, because the air pollution concentrations at participants' home and university stay at a very low level, and energy expenditure data of working and home activity is almost the same, we treat home activity and work equally -- which means we can place the original home activity period in the working period if there are some extra working hours based on different working hour constraints. We ran all evaluations on an Apple MacBook Pro with 2.7 GHz Intel Core i5 and 6 GB of DDR3 RAM.

\begin{table*}[!t] \renewcommand{\arraystretch}{1.1}
\newcommand{\tabincell}[2]{\begin{tabular}{@{}#1@{}}#2\end{tabular}}
  \caption{Scenario attributes}
   \centering
  \begin{tabular}{p{3.5cm} |p{6cm} | p{5cm}}
  \hline
      Commute pattern & Start to commute time window  & Working hours time window \\
    \hline
    Walking &   8:00 am -  9:00 am & 7 hours - 9 hours\\
    Cycling &   8:00 am -  10:00 am & 7 hours - 10 hours\\
    Driving &   8:00 am -  9:00 am & 7 hours - 10 hours\\
  \hline
  \end{tabular}
  \begin{tabular}{p{7.5cm} p{7.5cm}}
      commuting duration from residence to work & commuting duration from work to residence\\
    \hline
       0.35 hour &   0.35 hour\\
       1 hour &   1 hour\\
       1.1 hours &   1.1 hours\\
  \hline
  \end{tabular}
    \label{table:tmcevaluation}
\end{table*}

\textbf{Schedule optimization}

\begin{figure*}[!t]
  \centering
  \subfigure[]{
    \includegraphics[scale=0.35]{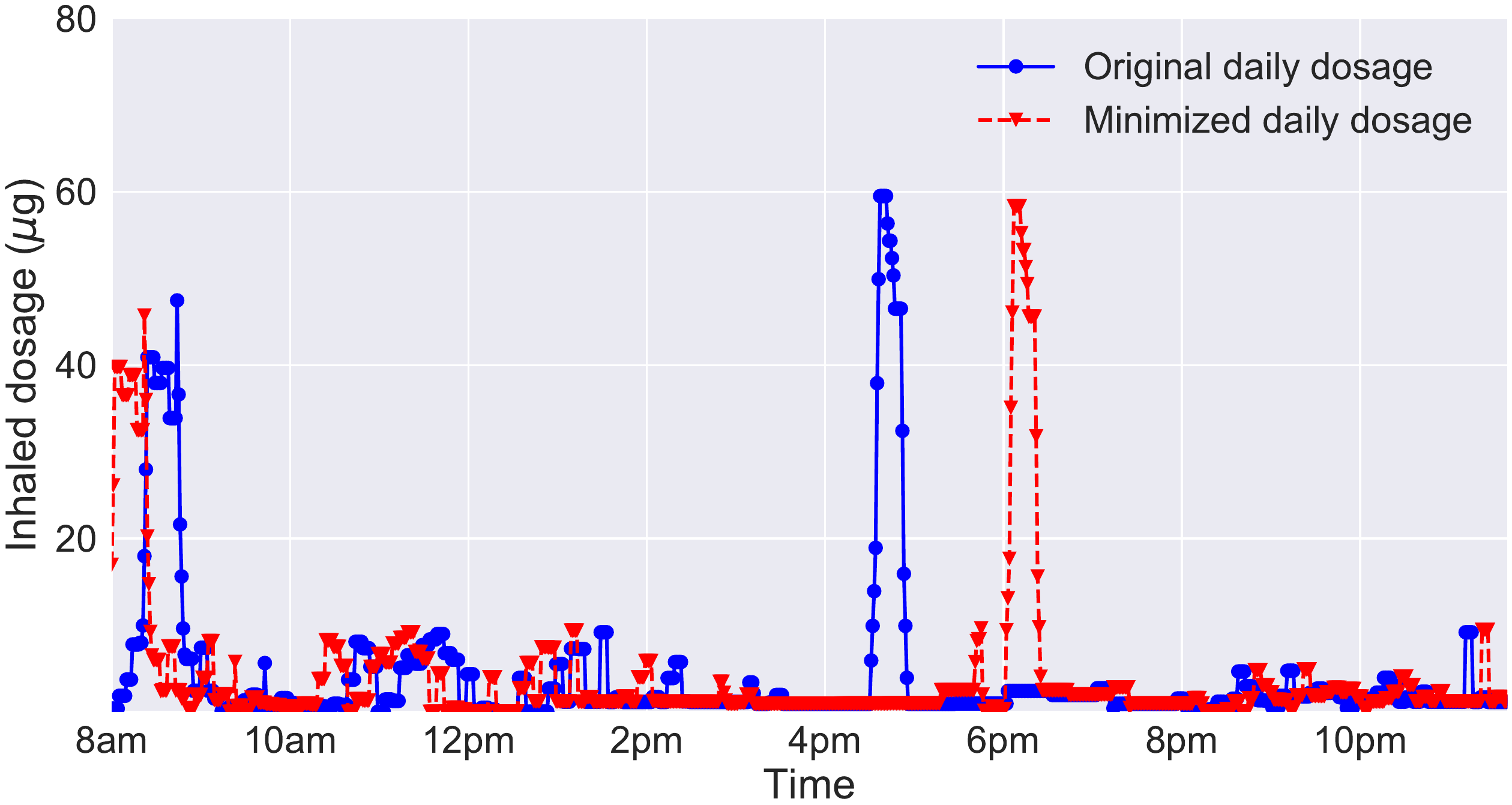}
      \label{fig:walkerwholedaydosage}}
        \quad
    \subfigure[]{
    \includegraphics[scale=0.35]{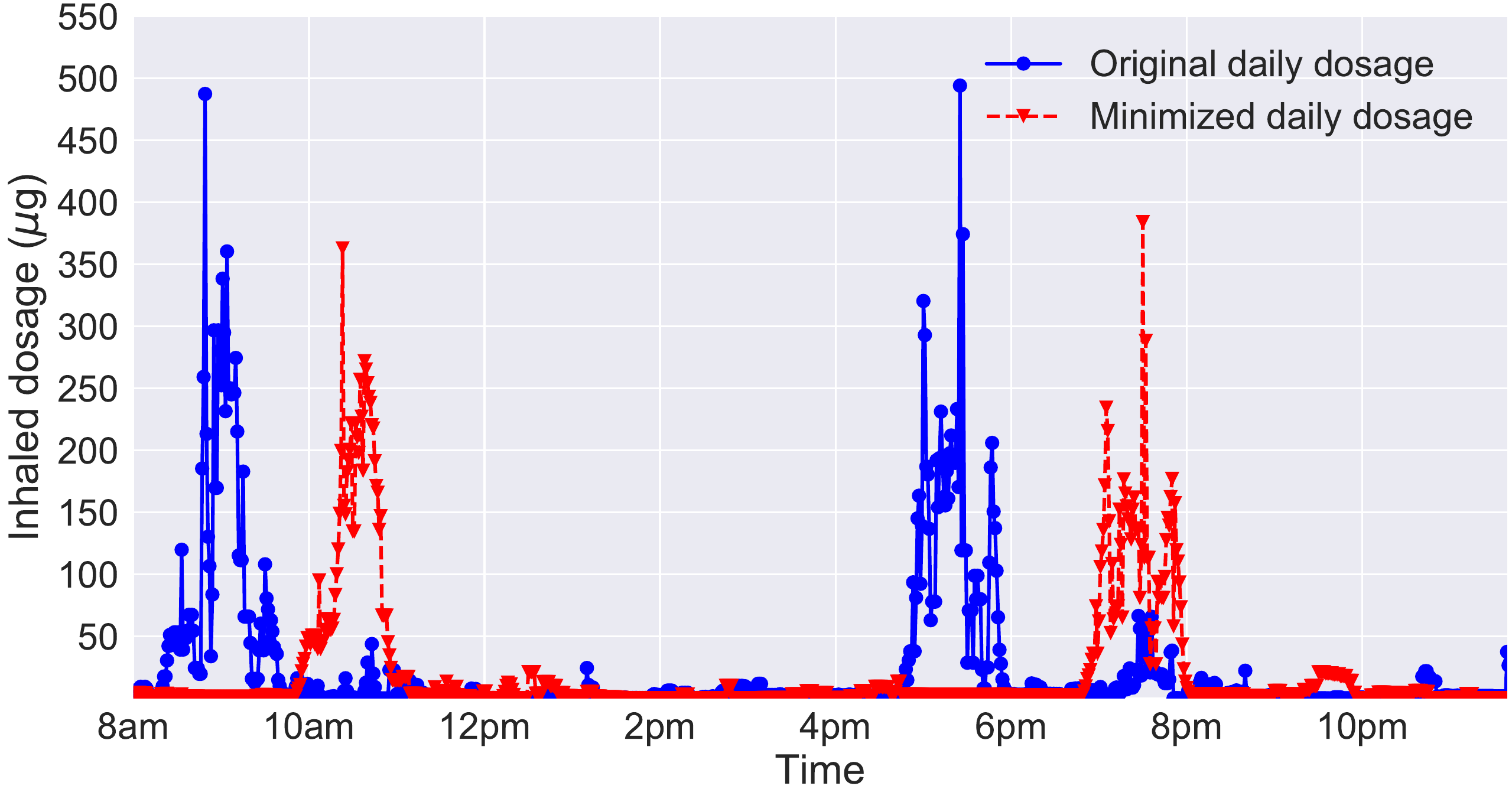}
      \label{fig:bikerwholedaydosage}}
        \quad 
    \subfigure[]{   
    \includegraphics[scale=0.35]{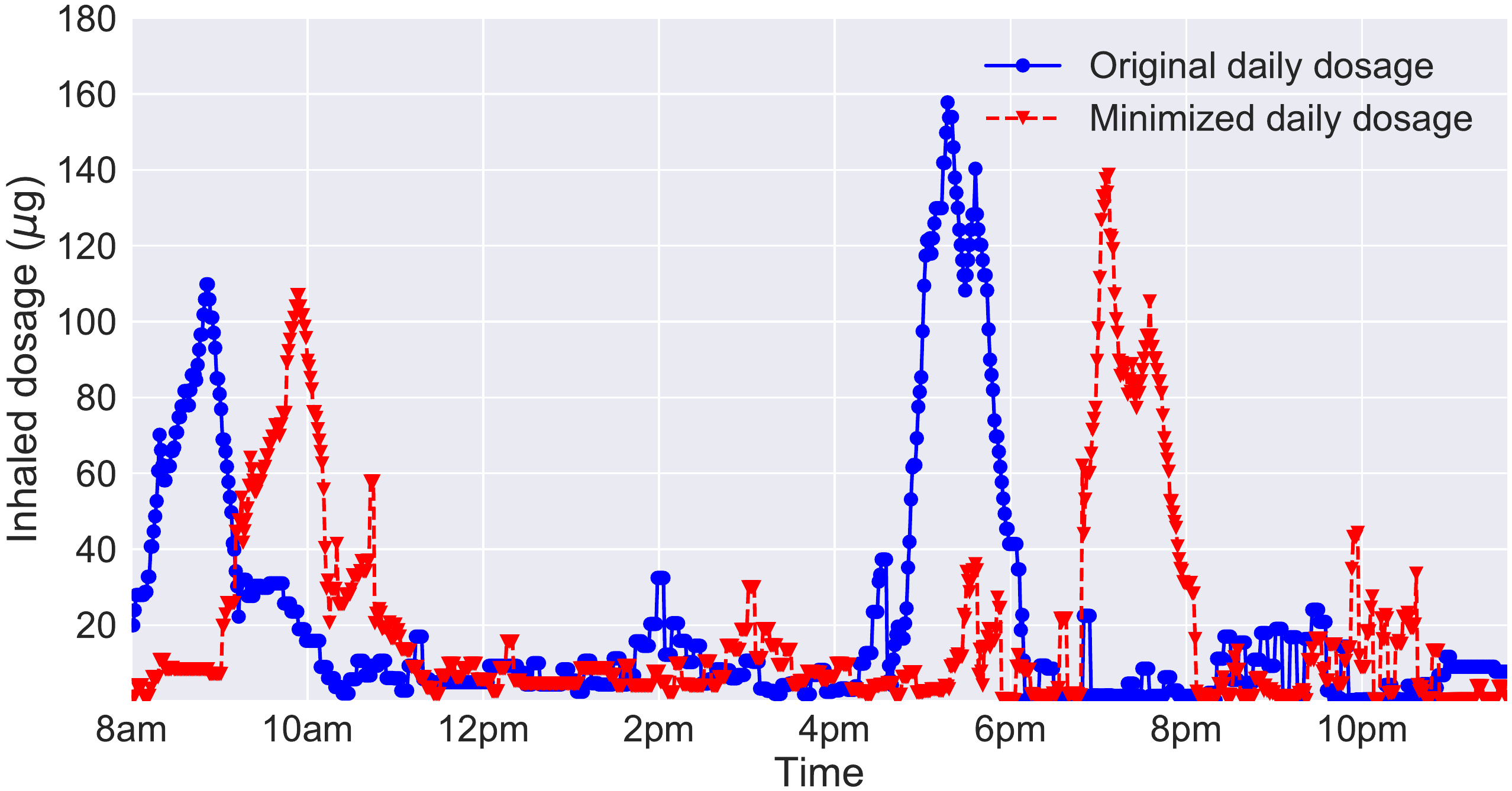}
      \label{fig:driverwholedaydosage}}
        \quad    
    \caption{Original and minimized daily inhaled dose by (a) one walker, (b) one biker, and (c) one driver.}
    \label{fig:wholedaydosagecompare}
\end{figure*}

We first evaluated our brute-force approach assuming the commuting routes of the three scenarios are fixed, and calculate the minimum inhaled dosage for each scenario based on our minimization algorithm. The three scenarios and constraints are shown in Table \ref{table:tmcevaluation},  After adopting our minimization algorithm and solution algorithm, the original and minimized results are shown in Fig. \ref{fig:wholedaydosagecompare}. Two direct observation can be made: 1) Personal daily dosage only peaks during commuting, no matter what the commute patterns are. All the other activities such as walking, working, and home activities do not result in significant inhaled dosage; 2) Our minimization algorithm works fine for the Biker and Driver, but not so well for the Walker. The reason is that the Walker lives near campus, where the pollution concentrations do not change much from his home to university. Even if we change the commute and working hours time window, the inhaled dosage will not change much. We can also see that the execution time for Biker is the longest, and it is almost three times the length of Walker and two times the length of Driver. The reason is that the longer start to commute time window and longer working hours for the Biker enlarge the searching space, and the one hour commuting duration also increases the number of calculations.

Table \ref{table:tmcdosagereduction} shows the whole day dosage inhaled by the Walker, Biker, and Driver using original and optimized commuting time. It can be seen that our minimization model achieves a high 14.1\% daily dosage reduction for the Biker, and 8.4\% for the Driver. We can also see that the Biker's daily inhaled dosage is a bit more than the driver, and almost six times higher than the Walker before minimization. The situation will change after adopting the minimization algorithm: the Biker still inhaled the most, but the value is very close to the Driver's dosage.

\begin{table*}[!t] \renewcommand{\arraystretch}{1.1}
\newcommand{\tabincell}[2]{\begin{tabular}{@{}#1@{}}#2\end{tabular}}
  \caption{Dosage reduction based on fixed routes}
   \centering
  \begin{tabular}{p{1cm} p{3cm} p{4cm} p{4cm} p{1.5cm}}
  \hline
        &  Execution time (seconds)&Original daily dosage ($\mu$g)  & Minimized daily dosage ($\mu$g) & Reduction\\
    \hline
     Walker & 56 &  3614.8& 3593.6 & 0.6\%\\
     Biker &  178&20508.8 & 17617.9 & 14.1\%\\
     Driver & 94&19189.1 & 17575.7 & 8.4\%\\
  \hline
  \end{tabular}
    \label{table:tmcdosagereduction}
\end{table*}

\textbf{Joint schedule and route optimization}

In the schedule optimization, we assumed that the commute routes are fixed for both ways. However, this is not always the case in real life. For example, commuters will try to avoid the motorway which has an accident, and pick up another faster route from Google Maps. So we include our route detection pragramme which gets two to three alternative routes from Google Map API in the evaluation and investigate how our algorithm works on the joint schedule and route optimization. We still use the three scenarios and constraints which are shown in Table \ref{table:tmcevaluation} for result comparison. Similar to the previous part, the whole day dosage inhaled by the Walker, Biker, and Driver using original and optimized commuting time are shown in Table \ref{table:tmcdosagereduction1}. Comparing with the results shown in Table \ref{table:tmcdosagereduction}, three immediate observations can be made: 

1) Execution time for Walker, Biker, and Driver are all increased greatly, almost nine times the length of the previous training time. The reason why the execution time increased massively is that adding alternative routes to the algorithm enlarge the calculation complexity. If the average route number is $R_{i}$ and $R_{j}$ for home to work and back home respectively, the calculation complexity will increase by $R_{i} \times R_{j}$ times.

2) Dosage is reduced even further for Biker and Driver, while Walker stays at the same reduction rate. The huge further dosage reduction for Biker and Driver indicates that alternative routes can reduce the dosage a lot, and including joint routes into minimization algorithm can enhance the model performance. The reason for the same dosage reduction for Walker is that the air pollution concentrations around university and pavement are very low, and walking time affects dosage the most.

\begin{table*}[!t] \renewcommand{\arraystretch}{1.1}
\newcommand{\tabincell}[2]{\begin{tabular}{@{}#1@{}}#2\end{tabular}}
  \caption{Dosage reduction based on alternative routes}
   \centering
  \begin{tabular}{p{1cm} p{3cm} p{4cm} p{4cm} p{1.5cm}}
  \hline
        &  Execution time (seconds)&Original daily dosage ($\mu$g)  & Minimized daily dosage ($\mu$g) & Reduction\\
    \hline
     Walker & 498 &  3614.8& 3593.6 & 0.6\%\\
     Biker &  1421&20508.8 & 16265.9 & 20.7\%\\
     Driver & 717 &19189.1 & 14214.7 & 25.9\%\\
  \hline
  \end{tabular}
    \label{table:tmcdosagereduction1}
\end{table*}

3) Comparing with the previous results which are shown in Table \ref{table:tmcdosagereduction}, we can see that Driver has more dosage reduction than Biker when including alternative routes. The reason is that the Driver drives along the road and motorway where the pollution concentrations are highest, and Driver's dosage is more sensitive to the routes than the Biker who is farther away from the highway and tunnel, typically on side road. For example, if an alternative route can avoid motorway and tunnel, the pollution concentrations along the route can be highly reduced, leading to a lower inhalation dosage for Driver. 

\subsection{Evaluation of the Heuristic Solution}\label{sec:tmcheuristic}

\begin{figure*}[!t]
  \centering
    \includegraphics[scale=0.55]{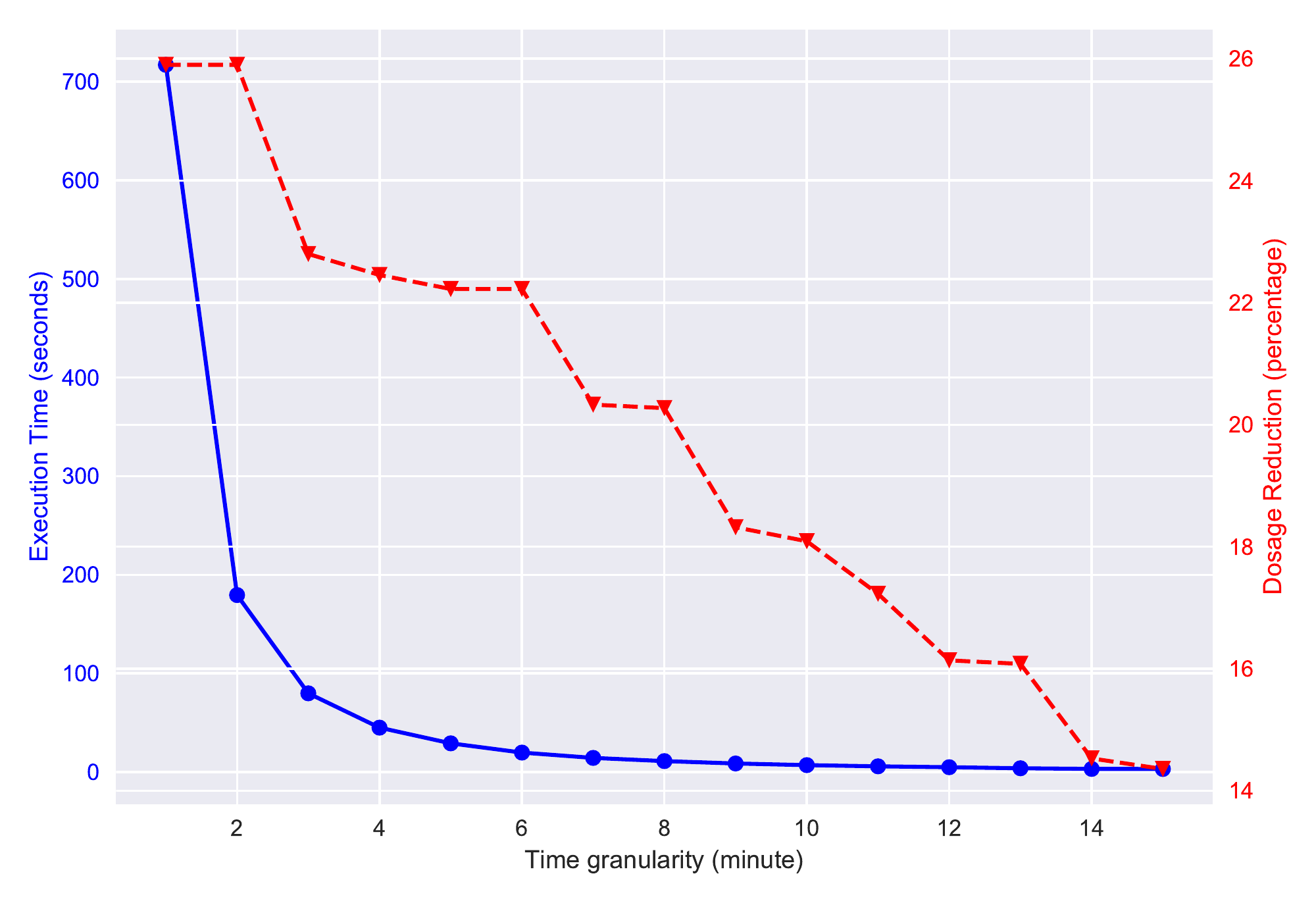}
    \caption{Comparison of execution time and dosage reduction based on different time granularity}
    \label{fig:executiontime}
\end{figure*}

Although including joint routes along with the time schedule indeed optimize the algorithm performance, the longer execution time is a big concern. To reduce the execution time while keeping a certain dosage reduction rate, we use the heuristic algorithm, which uses different time granularity $t$ from the current setting one minute to fifteen minutes which is one eighth of the departure time window. We use the Driver scenario as an example because it is more sensitive to the alternative routes. The result is shown in Fig. \ref{fig:executiontime}, from which we can see that increasing the time granularity significantly reduced the execution time, which decreases from 717 seconds to 11 seconds when the time granularity rises from one minute to eight minutes. The increasing time granularity also brings down the dosage reduction percentage, but less significantly compared with the execution time. It only dropped from 25.9\% to 20.3\% during the same period.

\begin{table*}[!t] \renewcommand{\arraystretch}{1.1}
\newcommand{\tabincell}[2]{\begin{tabular}{@{}#1@{}}#2\end{tabular}}
  \caption{Dosage reduction based on eight minutes time granularity}
   \centering
  \begin{tabular}{p{1cm} p{3cm} p{4cm} p{4cm} p{1.5cm}}
  \hline
        &  Execution time (seconds)&Original daily dosage ($\mu$g)  & Minimized daily dosage ($\mu$g) & Reduction\\
    \hline
     Walker & 7.2 &  3614.8& 3593.6 & 0.6\%\\
     Biker &  22.1 &20508.8 & 16937.2 & 17.4\%\\
     Driver & 10.9 &19189.1 &  15293.7& 20.3\%\\
  \hline
  \end{tabular}
    \label{table:tmcdosagereduction2}
\end{table*}

To balance the minimization result and execution time, a certain time granularity needs to be set. While the time granularity is a judgmental parameter, we set it to eight minutes to reduce the algorithm execution time while still achieve acceptable dosage reduction rate (higher than 20\% for Driver). The updated dosage reduction percentage for three scenarios is shown in Table \ref{table:tmcdosagereduction2}. We can see that the execution time has been reduced extensively for each scenario. The execution time is even lower than the fixed route result which has been shown in Table \ref{table:tmcdosagereduction}, while the dosage reduction percentage for Biker and Driver increased by 3.3\% and 11.9\% respectively. Comparing with the global optimal which has been shown in Table \ref{table:tmcdosagereduction1}, the reduction rate for Biker and Driver has increased by 3.3\% and 5.6\% respectively, while the execution time almost reduced to one seventieth.

\section{Conclusion and Future Work}\label{sec:tmcconclusion}
In this paper, we developed HazeDose system, which can personalize the air pollution exposure by individuals. Specifically, we combine the pollution concentrations obtained from the HazeEst system with the activity data from the individual's on-body activity monitors to estimate the personal inhalation dosage of air pollution. Users can visualize their personalized air pollution exposure information via a mobile application. We show that different activities, such as walking, cycling, or driving, impacts their dosage, and commuting patterns contribute to a significant proportion of an individual's daily air pollution dosage. Moreover, we propose a dosage minimization algorithm, with the trial results showing that up to 14.1\% of a biker's daily exposure can be reduced applying our algorithm using fixed routes, while using alternative routes the Driver can inhale 25.9\% less than usual. One heuristic algorithm is also introduced to balance the execution time and dosage reduction for alternative routes scenarios. The results show that up to 20.3\% dosage reduction can be achieved when the execution time is almost one seventieth of the original one.

We acknowledge that we only consider the dosage minimization for working days in this paper. However, the joint schedule and route options for weekend scenario are more complicated. The schedule flexibility and multiple origin/destination pairs bring more calculation complexity and some other heuristic or stochastic methods need to be included to reduce the algorithm execution time. As future work, we would like to include weekend scenario in the dosage minimization algorithm and we believe the HazeDose system is a step towards enabling accurate medical inferencing of the impact of long-term air pollution on individual health.

\section*{References}

\bibliography{HazeDose}

\end{document}